\documentstyle[epsfig,rotating]{mn}

\def\spose#1{\hbox to 0pt{#1\hss}}
\def\lta{\mathrel{\spose{\lower 3pt\hbox{$\mathchar"218$}}
     \raise 2.0pt\hbox{$\mathchar"13C$}}}
\def\gta{\mathrel{\spose{\lower 3pt\hbox{$\mathchar"218$}}
     \raise 2.0pt\hbox{$\mathchar"13E$}}}

\title
[The Off State of GX\,339--4]
{The Off State of GX\,339--4}
\author[A.K.H. Kong, E. Kuulkers, P.A. Charles and L. Homer]
{A.K.H. Kong$^1$\thanks{Email: albertk@astro.ox.ac.uk}, 
 E. Kuulkers$^{2,3}$, 
 P.A. Charles$^1$ and 
 L. Homer$^1$ \\
$^1$ Department of Astrophysics, Nuclear \& Astrophysics
Laboratory, Keble Road, Oxford OX1 3RH \\
$^2$ Space Research Organization Netherlands, Sorbonnelaan 2, 3584 CA
Utrecht, the Netherlands \\
$^3$ Astronomical Institute, Utrecht University, P.O. Box 80000, 3507 TA
Utrecht, the Netherlands \\
}
\date{Accepted. Received.}

\begin{document}

\maketitle

\begin{abstract}
We report {\it BeppoSAX} and optical observations of the black hole
candidate GX\,339--4 during its X-ray `off' state in 1999. The
broad-band
(0.8--50 keV) X-ray emission can be fitted by a single power law with
spectral index, $\alpha \sim 1.6$. The observed luminosity is
$6.6 \times 10^{33}$ erg s$^{-1}$ in the 0.5--10 keV band, which is
at the higher end of the flux distribution of black
hole soft X-ray transients in quiescence, comparable to
that seen in GS\,2023+338 and 4U\,1630--47. An optical
observation just before the {\it BeppoSAX} observation shows
the source to be very faint at these wavelengths as well ($B=20.1$,
$V=19.2$). By comparing with previously
reported `off' and low states (LS), we conclude that the `off' state
is actually an extension of the LS, i.e. a LS at lower intensities. We
propose that accretion models such as the
advection-dominated accretion flows are able to explain the
observed properties in such a state.
\end{abstract}


\section{Introduction}

The black hole candidate GX\,339--4 was discovered by Markert et al. (1973)
with the OSO--7 satellite and was soon noted for its similarity in
X-rays to the classical black hole candidate Cyg\,X--1 (Market et al.
1973; Maejima et al. 1984; Dolan et al. 1987). The source exhibits
aperiodic and quasi-periodic modulations on time scales spanning from
milliseconds to years over a wide range of wavelengths. It spends most of the
time in the so-called X-ray low state (LS) which has a power-law spectrum with
spectral index $\alpha \sim 1.5-2$ (Ricketts 1983; Maejima et al. 1984) and
strong (30--40\% rms) band-limited noise (Nowak et al. 1999; Belloni et al.
1999). In the high
state (HS), it becomes brighter (in the 2--10 keV band) and exhibits an
ultra-soft spectral component plus a steeper power-law (Maejima et al. 1984;
Belloni et al. 1999), while the temporal variability is only a few
percent
rms (Grebenev et al. 1993; Belloni et al. 1999). It also shows a very
high state (VHS; Miyamoto et al. 1991) with broad band noise of 1--15\% rms
and 3--10 Hz
quasi-periodic oscillations (QPOs) seen in its fast time variability, but
with a higher X-ray luminosity than in the HS.
Recently, an intermediate state (IS) was reported by M\'endez and van der Klis
(1997) and its spectral and timing properties are similar to the VHS but with a
much lower luminosity. Finally, an `off' state has also been reported
(see Markert et al. 1973; Motch et al. 1985; Ilovaisky et al. 1986;
Asai et al. 1998), in which the X-ray fast time variability is consistent
with that seen in the LS (M\'endez \& van der Klis 1997) while the
energy spectrum (power law with $\alpha$ of 1.5--2) is similar to the LS but
with a 2--10keV flux which
is $\sim 10$ times lower or even fainter than in the LS. It has
already been suspected that the `off'
state is in fact a weak LS (see e.g. van der Klis 1995). A summary of
the different states and their properties is given in Table 1. 

The optical counterpart of GX\,339--4 was identified by Doxsey et al. (1979)
as a $V\sim 18$ blue star, but subsequent observations showed that it exhibited
a wide range of variability from 
$V=15.4$ to 20.2 (Motch et al. 1985; Corbet et al. 1987) in its X-ray LS
and 
`off' state, while $V=16-18$ (Motch et al. 1985) in the X-ray HS.
Simultaneous
optical/X-ray observations also showed a remarkable anti-correlation in the
optical
and soft X-ray (3--6 keV) fluxes during a transition from X-ray LS to
HS (Motch et al. 1985), the cause of which is unknown. However, Ilovaisky et
al. (1986) showed that there are times when the optical
flux can be correlated with the X-ray luminosity. A possible
orbital period of 14.8 hr from optical photometry was
reported by Callanan et al. (1992). At present, there is no dynamical
mass estimate available for the compact object (which would establish
the black-hole nature of the compact object), since there has not yet
been a spectroscopic detection of the mass-losing star.

In this Letter, we report on recent {\it BeppoSAX} and optical
observations of GX\,339--4 during its current X-ray `off' state and compare
these data
with black hole soft X-ray transients (BHSXTs) in quiescence.            

\begin{table*}
\begin{minipage}{170mm}
\footnotesize
\caption{Comparison of different X-ray states of GX\,339--4.}
\begin{tabular}{l l c c c}
\hline
X-ray state& Energy spectrum& Power spectrum& Flux 2--10 keV$^a$& References\\
\hline
`Off'& power law (PL), $\alpha\sim 1.5-2$& $<$ 26\% rms& $<$ 1& this work,
1--5\\
Low (LS)& PL, $\alpha\sim1.5-2$& $\sim 25-50\%$ rms&
1--13& 6, 7\\
Intermediate (IS)& ultra-soft (US) component + PL, $\alpha \sim 3.5$& 7 \%   
rms& 15& 5\\
High (HS)& US component + weak PL, $\alpha\sim2-3$&
few \% rms& 70--80& 7, 8\\
Very high (VHS)& US component + PL, $\alpha \sim2.5$& 1--15\% rms, 3--10 Hz
QPO& 200& 9\\
\hline
\end{tabular}
$^a$ in units of $10^{-10}$ erg cm$^{-2}$ s$^{-1}$\\
References: (1) Nolan et al.
1982; (2) Motch et al. 1985; (3) Ilovaisky et al.
1986; (4) Asai et al. 1998; (5) M\'endez \& van der Klis 1997;
(6) Wilms et al. 1999; (7) Belloni et al.
1999; (8) Makishima et al. 1986; (9) Miyamoto et al. 1991
\end{minipage}
\end{table*}
\normalsize

\section{Observations and Data Reductions}

\subsection{{\it RXTE}/ASM}

The All Sky Monitor (ASM; Levine et al. 1996) on board the {\it Rossi
X-ray Timing Explorer} ({\it RXTE}; Bradt et al. 1993) has
monitored GX\,339--4 several times daily in its 2--12 keV pass-band
since February 1996.
The source remained in a low flux level ($\sim 2$ ASM cts/s) 
until early January 1998 ($\sim$ MJD 50810) although some variations were
seen (see Fig. 1). Pointed Proportional Counter Array (PCA)
observations during this period indicate that it is in the LS
(Wilms et al. 1999). After MJD 50800, the source flux increased dramatically
to $\sim 20$ ASM cts/s where it stayed for nearly 200 days before declining.
Belloni et
al. (1999) reported that the source underwent a LS to HS transition, probably
through an IS ($\sim$ MJD 50820). The source changed back
to the LS again in February 1999 ($\sim$ MJD 51200) as indicated by a sharp
increment in the
hard X-rays (BATSE: 20--100 keV) and radio emission (Fender et al. 1999).
Note that the ASM hardness ratio (5--12/1.3--3 keV) rises
significantly (see Fig. 1, lower panel) when the source changes from
HS to LS.
After June 1999 ($\sim$ MJD 51330), the ASM count rate dropped further and the
source intensity fell below the 3-$\sigma$ detection level (see Fig.
1). This is a strong indication that the source entered a so-called
`off' state at that time.

\subsection{{\it BeppoSAX} NFI}

We observed GX\,339--4 with the Narrow Field Instruments (NFI) on board {\it
BeppoSAX} between August 13.5 and 14.1, 1999
UT (marked in Fig. 1). The NFI consist of two
co-aligned imaging instruments providing 
a field of view of $37' \times 57'$: the
Low-Energy Concentrator Spectrometer (LECS; 0.1--10 keV; Parmar et al.
1997) and the Medium Energy Concentrator Spectrometer (MECS; 1.6--10.5 keV;  
Boella et al. 1997).  The other two NFI, non-imaging instruments are
the Phoswich Detector System (PDS; 12--300keV; Frontera et al. 1997) and
the High-Pressure Gas Scintillation Proportional Counter (HP--GSPC; 4--120 
keV; Manzo et al. 1997).

During our observations, the HP--GSPC was turned off due to
its recent anomalous behaviour (see the news web page of the {\it
BeppoSAX} SDC\footnote{http://www.sdc.asi.it/latestnews.html}). The net
exposure times are 12.8 ks for the LECS, 24.6 ks for the MECS
and 11.1 ks for the PDS. The J2000 coordinates of the source derived
from the MECS data are
R.A.=17$^h$ 02$^m$ 48$^s$, Dec.=-48$^{\circ}$ 47$'$ 37.5$''$, with a
90\%
confidence uncertainty radius of 56$''$ (see Fiore et al. 1999) which
is consistent with the position
of GX\,339--4. We confirm that there is no other X-ray source in
the field of view which could potentially contaminate our
target. We applied an extraction radius of $4'$ centred on 
the source position for both LECS
and MECS images so as to obtain the source lightcurves and spectra. The MECS
background was extracted by using long archival exposures on empty sky
fields. We also checked the background of the source-free regions
in the image and it is similar to the empty sky fields.
For the LECS spectrum, we extracted the background from two
semi-annuli in the same field of view as the source (see Parmar et al.
1999 for the reduction procedure). Both the
extracted spectra were rebinned by a factor of 3 so as to accumulate at least
20 photons per
channel and to sample the spectral full-width at half-maximum resolution (Fiore
et al. 1999). A systematic
error of 1\% was added to both LECS and MECS spectra to take account
of the systematic uncertainties in the detector calibrations (Guainazzi et
al. 1998). 
Data were selected in the energy ranges 0.8--4.0 keV (LECS),
1.8--10.5 keV (MECS) and 15--220 keV (PDS) to ensure a better
instrumental calibration (Fiore et al. 1999). A normalization factor was
included for the LECS and PDS relative to the MECS in order to correct
for the NFIs' flux intercalibration (see Fiore et al. 1999).

\subsection{Optical}

The optical counterpart of GX\,339--4 (V821 Ara) was observed at the South
African Astronomical Observatory (SAAO)
using the 1.9-m telescope and the UCT-CCD fast photometer
(O'Donoghue 1995) on 1999 August 10 (i.e. 3 days before the {\it BeppoSAX}
observation reported here), when the soft X-ray (2--12 keV)
flux was very low (see also Fig. 1). The observing conditions were
generally good
with typical seeing $\sim$ 1.5 arcsec. The exposure times on GX\,339--4 were
240s in the $B$-band and 60s in the $V$-band. Debiasing and flat-fielding were
performed with standard {\tt IRAF} routines.
Due to moderate crowding of the counterpart with a nearby but fainter
neighbour, point spread function (PSF) fitting was employed in order to
obtain good photometry with {\tt DAOPHOT II} (Stetson 1987). The
instrumental magnitude of our target star was then calibrated into the 
standard UBV system using a standard star of similar colour and observed at
approximately the same period. Since the standard star counts were determined
using a large aperture, local standard stars in the target field were used to
determine the offset between the PSF magnitudes and these aperture results.

\section{Results}

During both the {\it BeppoSAX} NFI and optical observations, the source
is barely
detected in the {\it RXTE}/ASM and the count rate often drops below the
3-$\sigma$ ASM
detection limit. The lightcurve and hardness ratio shown in Fig. 1 suggest
that the source transited from the LS
to a lower flux level, presumably an `off' state after MJD 51330. The
{\it BeppoSAX}
lightcurves in the various energy bands do not show any evidence 
for variability on timescales from 100s to 
5000s (the 3-$\sigma$ upper limit on the semi-amplitude is 0.3\%).  We have
also checked for
the LS-like fast time ($<$ 100 s) variability as seen typically in the
LS of black hole X-ray
binaries (see e.g. van der Klis 1995), but low counting statistics prevented
us from setting useful upper limits. 

The broad-band (0.8--50 keV) spectrum of GX\,339--4 from the
LECS, MECS and PDS data is satisfactorily ($\chi^2_{\nu}=1.09$ for 56
degrees of freedom (d.o.f.)) fitted by a single power-law plus absorption.   
The best fit spectral parameters are summarized in Table 2 and the
spectrum is shown in Fig. 3. We do not see any significant Fe-K 
line emission between
6.4--6.7 keV, with a 90\% confidence upper limit of $\sim 600$ eV on
the equivalent width. We note that there is a residual in
the LECS below 0.9 keV and this might be due to a very soft black-body
component or line emission near the Fe-L complex (e.g. Vrtilek et al.,
1988). We have also fitted the spectrum with single black-body and
bremsstrahlung models, but they are unacceptable ($\chi^2_\nu$ $>$ 2). 

In the optical, the source was seen at $B=20.1\pm0.1$ and 
$V=19.2\pm0.1$.

\begin{table}
\begin{minipage}{70mm}
\caption{Best-fit parameters for energy spectrum of GX\,339--4. The errors
are single parameter 1-$\sigma$ errors.}
\begin{tabular}{l c}
\hline
Model& Power law\\
\hline
$N_H$& $(5.1\pm1.5) \times 10^{21}$ cm$^{-2}$ \\
Photon index, $\alpha$& $1.64\pm0.13$ \\
$\chi^2_{\nu}$ (56 d.o.f.)& 1.09 \\
Flux$^a$ & $(2.2\pm0.2)
\times 10^{-12}$ erg cm$^{-2}$ s$^{-1}$ \\
$L_{X}$ $^b$ & $(6.6\pm0.4)\times 10^{33}$ erg s$^{-1}$
\\
\hline
\end{tabular}
$^a$ Absorbed flux in 2 -- 10 keV\\
$^b$ Luminosity in 0.5--10 keV, assuming 4 kpc (Zdiarski et al. 1998)\\
\end{minipage}
\end{table}

\section{Discussion}

Our X-ray (0.8--50 keV) and optical observations of GX\,339--4
took place
during a very low intensity (2--12 keV) X-ray state, presumably the
X-ray `off' state (see Fig. 1). Comparing with {\it
ASCA} and {\it RXTE} observations obtained when the source was in a LS (Wilms
et al. 1999; Belloni   
et al. 1999; see also Table 1), the spectral index and neutral hydrogen
column are similar ($\alpha \sim 1.6$, $N_H \sim 5\times 10^{21}$ cm$^{-2}$),  
but our observed flux of $2.2\times 10^{-12}$ erg cm$^{-2}$ s$^{-1}$ (2--10
keV) is much lower by 2--3 orders of magnitude. This confirms that the
source indeed changed to a very low luminosity state as indicated by the {\it
RXTE}/ASM data. Note that the observed column density is also
consistent with that derived from optical reddening, i.e.
$N_H=(6.0\pm0.6)\times10^{21}$ cm$^{-2}$ (see Zdziarski et al. 1998 for    
more detail). At a distance of 4 kpc, the observed
soft X-ray (0.5--10 keV) luminosity is $6.6\times10^{33}$ erg s$^{-1}$.
Ilovaisky et al. (1986) reported an `off' state seen by {\it EXOSAT} with a
luminosity of $1.1\times 10^{35}$ erg s$^{-1}$ in the 0.5--10 keV band; 
our measurement is a factor of $\sim$ 17 below that. We also note that an
upper limit was obtained
by {\it ASCA} in 1993 of $5\times 10^{32}$ erg s$^{-1}$ (Asai et al.
1998), which suggests that the source was in the `off' state as well.
In addition, Nolan et al. (1982) observed the source in the
12--200 keV band and claimed that one of the observations was in the `off'
state
with a flux level of $4.7\times 10^{-10}$ erg cm$^{-2}$ s$^{-1}$ in the
20--50 keV band.
This luminosity is only comparable with the recent {\it RXTE}/HEXTE
observations (Wilms et al. 1999) in the LS, while our PDS
observation indicates that the source was down to $\sim 3.5\times
10^{-12}$ erg cm$^{-2}$ s$^{-1}$ in the same energy band. Therefore, the Nolan
et al. (1982) observation was actually not in the `off' state. We
have obtained the {\it first} firm detection of GX\,339--4 at low
intensity up to 50 keV.

In the optical, our observed magnitude of $V=19.2$ and $B=20.1$ are faint
compared to the
wide range of reported magnitudes: $B \gta 21$ (Ilovaisky 1981) and
$V=15.4$ (Motch et al. 1982). Our result is, however, comparable to the
observations by Ilovaisky (1981) and Remillard and
McClintock (1987), and this is the third time that the $B$ magnitude has
been seen to fall to $\sim 20$. The observed colour, $B-V$ is 0.9, which is
also
consistent with previous observations in different X-ray states
(Makishima et al. 1986;
Ilovaisky et al. 1986; Corbet et al. 1987).
We note that the optical and X-ray emission can be anti-correlated
during state transitions (e.g. Motch et al. 1985; observations during
LS to HS transition). Our observations show that this may
not be the case in the current `off' state since both X-ray and optical
are at very low luminosity and it suggests they correlate
from the LS to `off' state.

More recently, Corbel et al. (1999) show that the X-ray and radio
fluxes during the `off' state and the LS are well correlated.    
GX\,339--4 was detected at a level of $0.27\pm0.06$ mJy (at 8640 MHz)
with the Australia Telescope Compact Array (ATCA) on
1999 August 17 (MJD 51407) and at a comparable level two weeks later
(Corbel et al. 1999). This is
one of the lowest levels ever detected in the radio, but still
considerably above the upper limits to the radio emission during the HS
(see Fender et al. 1999). It is very similar to GS\,2023+338 which
still showed
ongoing radio emission after the source was off (see Han \& Hjellming
1992). Adding all the results from the reported X-ray `off' state and LS,
and the behaviour at other wavelengths, it
can be concluded that the `off' state is indeed an extension of the LS,
but at lower intensity. One might 
therefore expect rapid ($<$ 100s) temporal variability in the `off'
state but we were unable to verify this due to the low counting statistics. We
note that M\'endez \& van der Klis (1997) derived a 3$\sigma$ upper limit on
the variability in the 0.002--10 Hz range of 26\% rms in their observations in
the `off' state.

\begin{table*}
\begin{minipage}{115mm}
\caption{Quiescent X-ray luminosities and luminosity ratio between the
X-rays and the optical of BHSXTs.}
\begin{tabular}{l c c c c c c}
\hline
Source& Photon & kT$^a$& Luminosity$^b$ & $L_X/L_{opt}$ $^c$& D &
References\\
      & index, $\alpha$& keV&($10^{33}$ erg s$^1$)& &(kpc)&\\
\hline
GS\,2023+338& 1.7& -& 2& 0.038& 3.5& 1, 2\\
4U\,1630--47& -& 0.2$^d$& 8& unknown& 10& 3\\
GX\,339--4& 1.6& -& 6.6& 0.27& 4& this work\\
A0620--00& 3.5& 0.16& 0.01& 0.07& 1& 2, 4\\
GRO\,J1655--40& 0.7& -& 0.25& 0.002& 3.2& 1, 2\\ 
\hline
\end{tabular}
$^a$ black-body temperature\\
$^b$ in 0.5--10 keV\\
$^c$ $L_{opt}$ in 300--700 nm\\
$^d$ assumed spectrum\\
References: (1) Asai et al. 1998; (2) Menou et al. 1999; (3) Parmar et al.
1997; (4) Narayan et al. 1997a\\
\end{minipage}
\end{table*}
\normalsize

Our observed luminosity in the 0.5--10 keV band is comparable to the
BHSXT GS\,2023+338 and 4U\,1630--47 in quiescence (Asai et al. 1998;
Menou et al. 1999; Parmar et al. 1997). The quiescent X-ray
luminosities and X-ray/optical luminosity ratios of several BHSXTs are
given in Table 3 for comparison. Note that the quiescent spectrum of
A0620--00 can be fitted either by a power law or black-body model,
presumably due to the narrow energy range of {\it ROSAT}. All the objects
except 4U\,1630--47 in
Table 3 have firm detections such that the spectra can be determined. We
note that BHSXTs can be fitted with a power law spectrum in general, while
neutron star SXTs (NSSXTs) can be fitted by power law or black-body models (see
Asai et al. 1998). Recently, Rutledge et al. (1999) fitted the spectra from
NSSXTs with a hydrogen atmosphere model and
found that the derived parameters (radius and kT) of A0620--00 and
GS\,2023+338 were different from those found for NSSXTs.
Although the results are
based on {\it ROSAT} data, Asai et al. (1998) show similar findings by
re-analysing {\it ASCA} data. It suggests that the quiescent
X-ray spectrum can provide additional information to distinguish between 
black holes and neutron stars. Significant
X-ray variability in quiescence was observed in GS\,2023+338
(Wagner et al. 1994), 4U\,1630--47 (Parmar et al. 1997) and A0620--00
(Asai et al. 1998; Menou et al. 1999), while we have obtained a similar
result for GX\,339--4 (i.e. by comparing with Asai et al. 1998 and other
`off' state observations). This is a strong
indication that the BHSXTs in quiescence are not totally turned off and
that the `off' state of GX\,339--4 is an extended LS, as discussed above.
GX\,339--4 is similar to the quiescent state of BHSXTs, as will also be
discussed below.

We convert the optical magnitude into an optical (300--700 nm) luminosity of
$2.4\times10^{34}$ erg s$^{-1}$ (assuming $A_V=3.6$ and $E_{B-V}=1.2$;
Zdziarski et al. 1998). The ratio of the soft X-ray (0.5--10 keV)
and optical (300--700 nm) luminosities, $L_X/L_{opt}$ is $\sim 0.27$,
which is higher than other BHSXTs (see Table 3). This could be due to a
somewhat higher
X-ray luminosity for GX\,339--4 (see Table 3). All
these results resemble the quiescent state spectrum predicted by
advection-dominated
accretion flow models (ADAF; see Narayan \& Yi 1995; Esin et al. 1997). In
the current ADAF model for BHSXTs in quiescence, the viscously dissipated
energy is stored
in the gas and advects into the black hole so as to account for the low
luminosity of BHSXTs in quiescence. This model is in good agreement with
the observed optical, UV and X-ray emission of systems such as
GS\,2023+338 and A0620--00 (Narayan et al. 1996; Narayan et al. 1997a).
More recently, Fender et al. (1999) and Wilms et al. (1999) have applied
ADAF models to explain the observed X-ray emission
of GX\,339--4 in the LS. Narayan et al. (1997b) point out that the ADAF
luminosity depends significantly on the accretion rate. Therefore 
GX\,339--4 maintains a certain accretion rate in the LS for most of
the time, but the `off' state reported here represents a sudden decrease in
the accretion rate. It therefore suggests that the source is not
totally turned off, as is observed.

Following Narayan et al. (1997), the mass accretion rate of GX\,339--4 in the
`off' state corresponds to $\sim 10^{16}$ g s$^{-1}$. It is also consistent
with the $\dot M \sim 10^{15}-10^{16}$ g s$^{-1}$ predicted by
binary-evolution models (King et al. 1996; Menou et al. 1999), assuming an
orbital period of 14.8 h (Callanan et al. 1992) and black hole mass of 5
$M_{\odot}$ (Zdiarski et al. 1998). This accretion rate is
very similar to that of GS\,2023+338 (Narayan et al. 1997a) in
quiescence. This relatively large accretion rate could be due to an evolved
companion star like GS\,2023+338 which supports the possibility of an
evolved subgiant in GX\,339--4 (Callanan et al. 1992). 
More detailed ADAF modelling may be needed in order to study the
nature of the `off' state of GX\,339--4 and make a direct comparison with
BHSXTs in quiescence.

Note that apart from the above, the observed range in $V$
magnitude of GX\,339--4 is similar to
the expected outburst amplitude for an SXT with an orbital
period of 14.8 hr (Shahbaz \& Kuulkers 1998). Together with the
evidence for the similarity of the spectrum and luminosity, we suggest
the `off' state of GX\,339--4 to be equivalent to BHSXTs in quiescence.
The result is important since GX\,339--4 undergoes state transitions on
much shorter timescales, while most of the BHSXTs exhibit outbursts
every 10--20 years. Finally, we note that we will need a good radial velocity
measurement of GX\,339--4 in its `off' state when the contamination of the
X-ray irradiated disc is at a minimum, so as to measure or constrain the black
hole mass.

\section*{Acknowledgments}
We are grateful to the {\it BeppoSAX} SDC science team for their
assistance in the data reduction. We thank Robert Rutledge and Lars
Bildsten for constructive comments. EK thanks Rob Fender for stimulating
discussions. This paper utilizies quick-look results provided by the
ASM/RXTE team. AKHK is supported by a Hong Kong Oxford Scholarship.

\bibliographystyle{}

\newpage

\begin{figure*}
{\rotatebox{-90}{\psfig{file=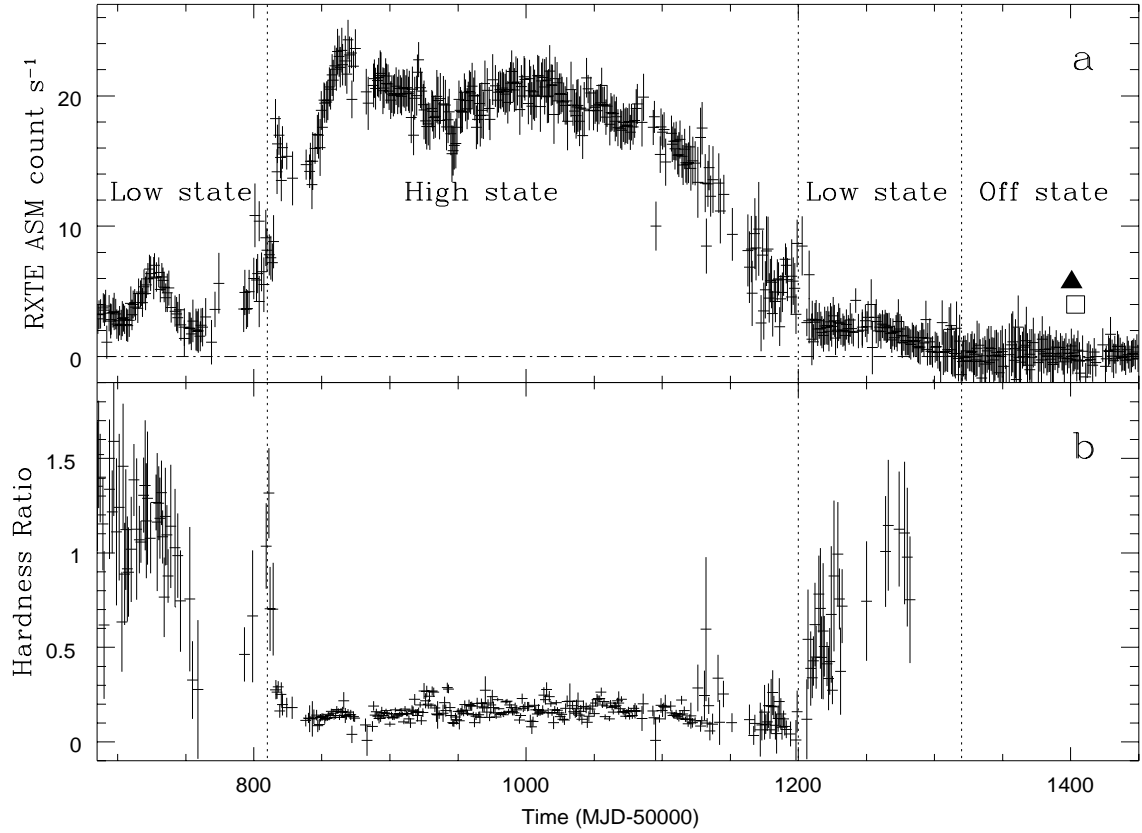,height=16.5cm,width=12cm}}}
\caption{\small{(a) {\it RXTE} ASM 2--12 keV light curve and (b) hardness
ratio (5--12/1.5--3 keV) curve of GX\,339--4 using one-day averages. This
clearly shows that the source has gone through
different state transitions during the last $\sim$ 3 years. The open
square marks the time of our {\it BeppoSAX} NFI
observations, while the filled triangle marks that of our optical
observation. The boundaries of the different states are mainly based on
Fender et al. (1999).}}
\end{figure*}

\newpage

\begin{figure*}
{\rotatebox{-90}{\psfig{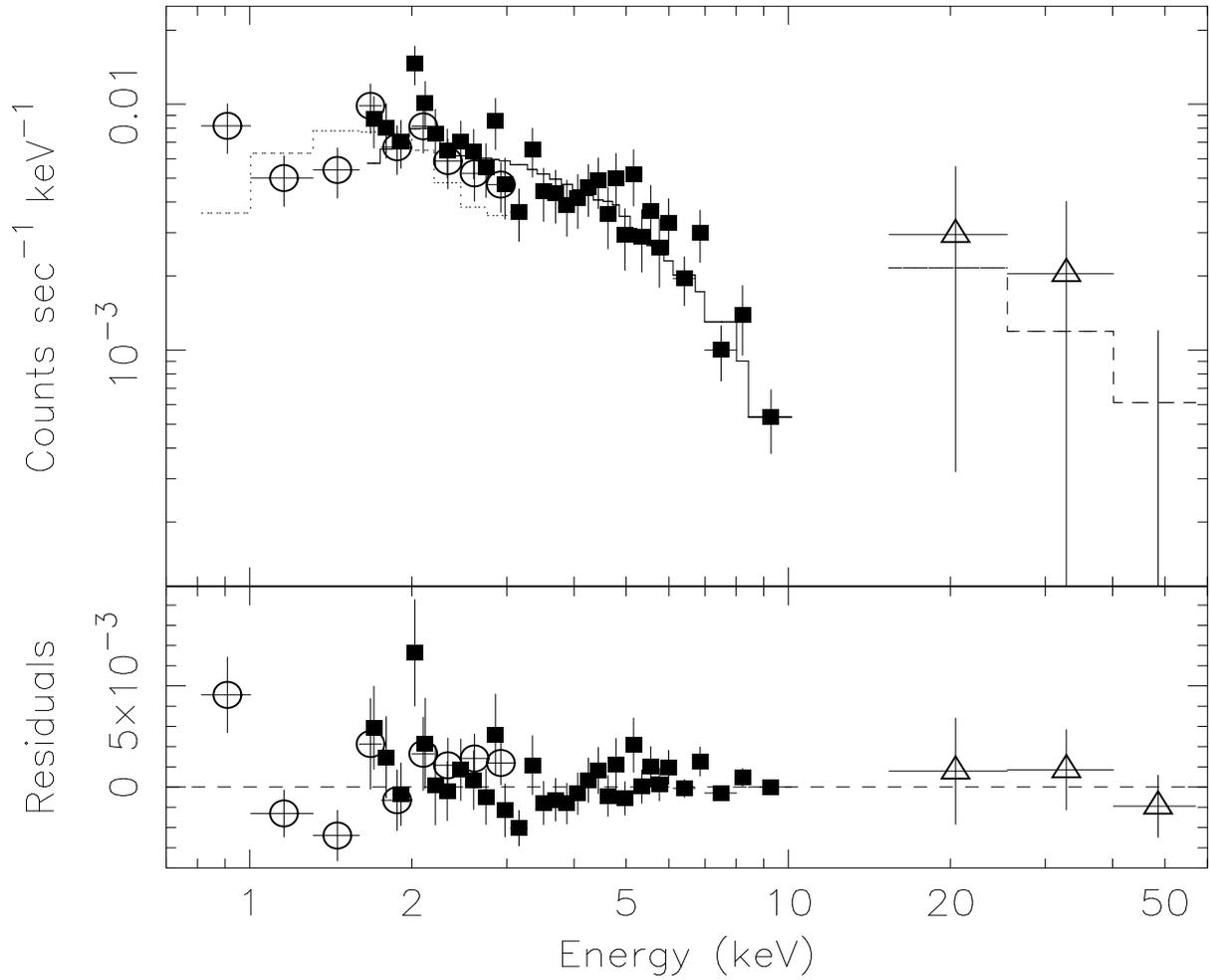}}}    
\caption{\small{Upper panel: {\it BeppoSAX} NFI spectral fit to GX\,339--4.
Open circles indicates LECS, solid square indicates MECS and triangles 
indicates PDS. The spectrum was fitted with a power-law  
($\alpha=1.64\pm0.13$). Lower panel: residuals after subtracting the fit
from the data in units of 1-$\sigma$.}}
\end{figure*}

\end{document}